\documentclass[10pt]{article}

\usepackage{amsmath}
\usepackage{amssymb}

\usepackage{graphicx}

\usepackage{cite}

\usepackage{color} 

\usepackage[labelfont=bf,labelsep=period,justification=raggedright]{caption}

\makeatletter
\renewcommand{\@biblabel}[1]{\quad#1.}
\makeatother

\date{}

\begin{document}

% Title must be 150 characters or less
\begin{flushleft}
{\Large
\textbf{Host mobility drives pathogen competition in spatially structured populations}
}
% Insert Author names, affiliations and corresponding author email.
\\
$\textrm{Chiara  Poletto}^{1,2,3,\ast}$, 
$\textrm{Sandro Meloni}^{4}$, 
$\textrm{Vittoria Colizza}^{2,3,5}$,
$\textrm{Yamir Moreno}^{4,6}$
$\textrm{Alessandro Vespignani}^{7}$
\\
\bf{1} Computational Epidemiology Laboratory, Institute for Scientific Interchange, Turin, Italy
\\
\bf{2} INSERM, U707, Paris, France
\\
\bf{3} UPMC Universit\'e Paris 06, Facult\'e de M\'edecine Pierre et Marie Curie, UMR S 707, Paris, France
\\
\bf{4} Institute for Biocomputation and Physics of Complex Systems, University of Zaragoza, Zaragoza, Spain
\\
\bf{5} Institute for Scientific Interchange, Turin, Italy
\\
\bf{6} Department of Theoretical Physics, University of Zaragoza, Zaragoza, Spain
\\
\bf{7} Laboratory for the Modeling of Biological and Socio-technical Systems, Northeastern University, Boston MA, USA
\\
$\ast$ E-mail: chiara.poletto@isi.it
\end{flushleft}

% Please keep the abstract between 250 and 300 words
\section*{Abstract}
Interactions among multiple infectious agents are increasingly recognized as
a fundamental issue in the understanding of key questions in public health, regarding pathogen emergence, maintenance, and evolution. The full description of host-multipathogen systems is however challenged by the multiplicity of factors affecting the interaction dynamics and the resulting competition that may occur at different scales, from the within-host scale to the spatial structure and mobility of the host population. Here we study the dynamics of two competing pathogens in a structured host population and assess the impact of the mobility pattern of hosts on the pathogen competition. We model the spatial structure of the host population in terms of a metapopulation network and focus on two strains imported locally in the system and having the same transmission potential but different infectious periods. We find different scenarios leading to competitive success of either one of the strain or to the codominance of both strains in the system. The dominance of the strain characterized by the shorter or longer infectious period depends exclusively on the structure of the population and on the the mobility of hosts across patches. The proposed modeling framework allows the integration of other relevant epidemiological, environmental and demographic factors opening the path to further mathematical and computational studies of the dynamics of multipathogen systems.

% Please keep the Author Summary between 150 and 200 words
% Use first person. PLoS ONE authors please skip this step. 
% Author Summary not valid for PLoS ONE submissions.   
\section*{Author Summary}

When multiple infectious agents circulate in a given population of hosts, they interact for the exploitation of susceptible hosts aimed at pathogen survival and maintenance. Such interaction is ruled by the combination of different mechanisms related to the biology of host-pathogen interaction, environmental conditions and host demography and behavior. We focus on pathogen competition and we investigate whether the mobility of hosts in a spatially structured environment can act as a selective driver for pathogen circulation. We use mathematical and computational models for disease transmission between hosts and for the mobility of hosts to study the competition between two pathogens providing each other full cross-immunity after infection. Depending on the rate of migration of hosts, competition results in the dominance of either one of the pathogens at the spatial level -- though the two infectious agents are characterized by the same invasion potential at the single population scale -- or cocirculation of both. These results highlight the importance of explicitly accounting for the spatial scale and for the different time scales involved (i.e. host mobility and spreading dynamics of the two pathogens) in the study of host-multipathogen systems.

\section*{Introduction}

While the dynamic of infectious diseases has been traditionally studied focusing on single pathogens one at a time, increasing attention is currently being devoted to the interactions among multiple infectious agents~\cite{Keeling2008}. Interaction mechanisms can indeed alter the pathogen ecology and have important evolutionary, immunological and epidemiological implications~\cite{Nowak1994,Rohani2003,Pedersen2007,Rohani2008}. 
A clear example of  pathogen cocirculation is given by viruses that may have different genetic and antigenic variants, such as human influenza A virus  with different subtypes and associated strains (i.e. phenotypically different variants)~\cite{Webster1992} and dengue virus with four serotypes circulating in affected tropical regions~\cite{Holmes2003}. Among other examples we find  many sexually transmitted diseases (like human immunodeficiency virus (HIV), human papilloma virus, herpes simplex virus), but also infections affecting animals, such as avian influenza~\cite{Webster1992} or the foot-and-mouth disease causing rapid acute infections in livestock~\cite{Haydon2001}.

The interaction among pathogens are mostly driven by  immune-mediated~\cite{Nowak1994} or ecological~\cite{Rohani2003} mechanisms, generally resulting into competition among the  infectious agents~\cite{Pedersen2007,Rohani2008}, even though cooperation may be observed in some specific settings~\cite{AbuRaddad2006}.
 Among strain-polymorphic  pathogens, for example, immune-mediated interaction occur when infection by a strain confers long-lasting protection against the particular strain, with partial cross-immunity against viral variants,  depending on the level of similarity of their genetic and antigenic profiles. Cocirculating strains are therefore not independent~\cite{Roberts2011}, as it happens in the case of influenza A virus, with strain-dependent prolonged immunity following infection~\cite{Webster1992} and epidemiological evidence for partial cross-immunity among strains~\cite{Sonoguchi1985,Sonoguchi1986}.
Interaction in terms of ecological interference is due to the temporary or permanent removal of a host from the population of susceptible hosts, because of infection from another strain. This may occur during the illness period and associated recovery (e.g. an individual staying at home or being admitted to the hospital) or because of deadly outcomes, generating complex competition dynamics for  the exploitation of the remaining hosts.

The spatial and social structure of a host population, as well as the migration of hosts, is recognized to represent a crucial element  affecting the geographical propagation of directly transmitted infectious diseases~\cite{Riley2007}. Infectious hosts moving from one location to another may seed the disease in previously unaffected locations, whereas susceptible hosts may contract the disease by entering in close contact with members of already infected subpopulations~\cite{McLean2005, Fraser2009, Balcan2009a, Keeling2005, Eubank2004}. Recently available massive datasets on host spatial structure and mobility patterns~\cite{Keeling2010,Bajardi2011b, Chowell2003, Barrat2004, Guimera2005, Brockmann2006, Gonzalez2008} have enabled the development of a large quantity of modeling approaches that assess the relevance and impact of hosts' mobility features on epidemic spreading processes caused by a single pathogen~\cite{Riley2007,Keeling2010,Colizza2006a, Balcan2009b, Grais2004, Cooper2006, Hufnagel2004, Chao2010, Ferguson2005, Merler2009, Colizza2007a, Colizza2007b, Colizza2008, Balcan2011, Balcan2012, Meloni2011, Poletto2012, Apolloni2013, Nicolaides2012}. Given its importance for dispersal mechanisms, the spatial structuring of the host population and the coupling among different subpopulations may also be  important to multipathogen dispersal mechanisms, and to epidemiological and ecological interactions~\cite{Galvani2003}. Space and host mobility may indeed act as an additional mechanism of ecological interference for host exploitation between different pathogens cocirculating in a population of susceptible hosts where the approximation of homogeneous mixing among hosts does not hold. Multistrain epidemics in the absence of homogeneous mixing have been studied assuming static networks or lattices, without considering host mobility~\cite{Funk2010, Newman2011, Marceau2011}, and in the context of pathogen evolution, often providing  detailed approaches regarding biological and epidemiological mechanisms (e.g. they properly account for pathogen mutation, physiological trade-off, cross-immunity and other relevant immunological and biological aspects) but lacking explicit modeling of host behavioral ecology regarding mobility~\cite{Haraguchi2000, Ballegooijen2004, Wild2009, Boots2010,Boerlijst2010}.

In view of all the elements at play in the study of host-multipathogen systems, a key question is therefore to assess to what extent patterns of coexistence or dominance of parasites are shaped by competition among infectious agents induced by specific mechanisms of interaction, as opposed to other biological factors characterizing individually the pathogens or the host population. In this paper we focus  on the competition mechanisms induced by hosts mobility in a spatially structured population in the case of two strains with full cross-immunity (an exploration of the partial cross-immunity case is reported in the Text S1). In order to single out the effect of mobility and population structure on the competition dynamic, we do not consider  pathogen evolution processes.  Furthermore we consider rapid acute infections, and ignore within-host interactions and within-host coexistence, which may instead be more relevant for persistent infections. This modeling framework represents a plausible setting for the analysis of a multistrain model for human influenza A in the framework of  a single epidemic season. In this case the immune-driven antigenic drift has been rarely observed in a geographically restricted region~\cite{Lavenu2006,Nelson2006}, suggesting that virus diversity is largely generated through importations instead of evolutionary mechanisms~\cite{Lavenu2006,Nelson2006,Nelson2008}. 

By introducing a general modeling framework in terms of a metapopulation approach, we find that changes in the host mobility rate alter the ecological conditions of the host-multipathogen system, which in turn changes the competitive balance between strains, resulting in a shift in their relative abundance and/or dominance. Given the importation of two strains with the same basic reproductive number (i.e. equal advantage  at the population level within each patch) but different timescales characterizing the infectious period, an increase in the host mobility selects the fast strain (i.e. the strain with the shorter infectious period) that becomes dominant in the system. On the contrary, fragmented population with low host mobility selects the slow strain (i.e. the strain with the longer infectious period) as it diffuses more efficiently  from one patch to another reaching the highest prevalence in the population. An intermediate mobility regime exists where the two strains codominate in the system.
Computational results are further supported by theoretical arguments. The simplifying assumptions considered in the model make it applicable to a large variety of host-multipathogen systems, as competition may arise in the interactions between strains but also of unrelated pathogens;  within this general framework we therefore use pathogen, parasite type, strain, or variant  as synonymous hereafter.

% You may title this section "Methods" or "Models". 
% "Models" is not a valid title for PLoS ONE authors. However, PLoS ONE
% authors may use "Analysis" 

%.................. METHODS

\section*{Methods}

\subsection*{Host-multipathogen infection model}

We consider a two-pathogen compartmental model that tracks hosts according to their pathogen-specific infection status. The infection by each strain is described by a susceptible-infectious-recovered (SIR) dynamics~\cite{Anderson1992}. We assume full cross-immunity, so that after infection by one strain the host is found to be fully immune to the other strain. We also assume that no other interaction among strains occurs besides full cross-immunity, therefore neglecting coinfection or superinfection events, a plausible assumption for rapid acute infections such as influenza. The case of partial cross-immunity among strains is also presented in the Text S1, however the full exploration of this case and of the resulting phase space of the system will be the object of further studies.

\begin{figure}[!ht]
\begin{center}
\includegraphics{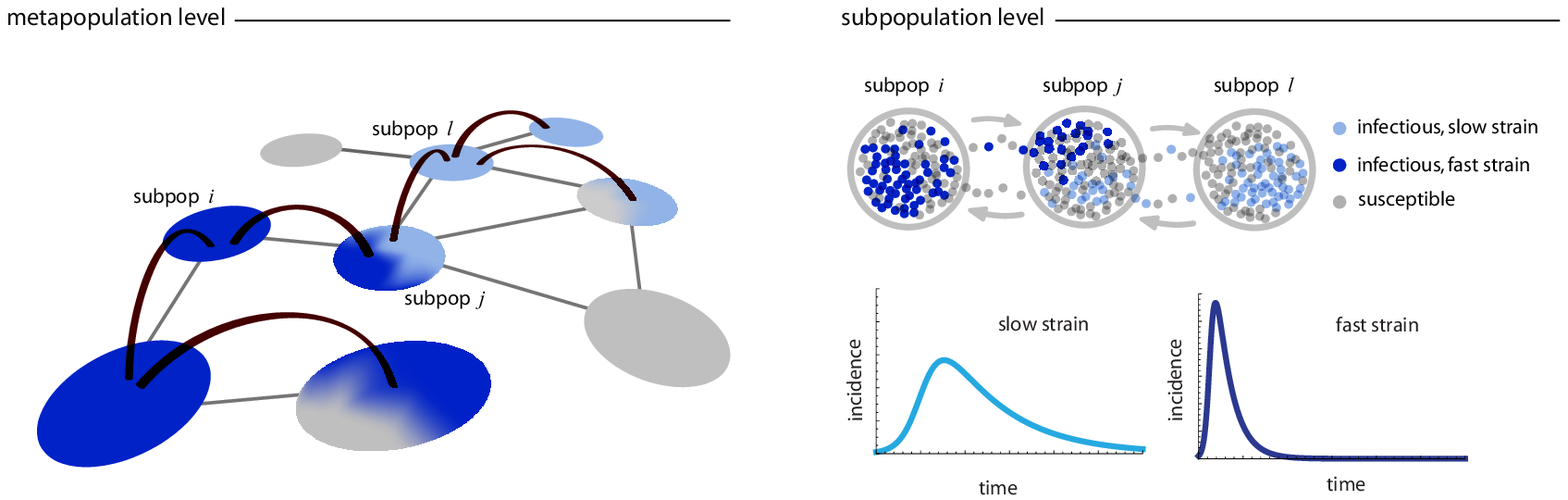}
\end{center}
\caption{
{\bf Schematic representation of the host-multipathogen metapopulation system.} At the macroscopic level the system is composed by a network of subpopulations connected via communication links that allow individuals to migrate from one subpopulation to the other. Inside each subpopulation the epidemic process take place. Susceptible individuals $S$ can be infected by the slow (fast) strain and change their status to $I_s\textrm{}(I_f$); infected individuals enter into the recovered class $R$ at rate $ \tau^{-1}\mu$ and $\mu$, for the slow and fast strain, respectively. Different epidemic waves are produced by the two strains when unfolding independently in a population, as shown by the number of new cases (incidence) over time.
}
\label{fig:model} 
\end{figure}

The SIR dynamics for strain $(i)$, with $i=\{1,2\}$, is ruled by the transition rates, $\beta_{(i)}$ and $\mu_{(i)}$, representing the disease transmissibility rate (for the transition from susceptible to infectious) and the recovery rate (for the transition from infectious to recovered), respectively. The dynamics is characterized by the basic reproductive number $R_0^{(i)}= \beta_{(i)}/\mu_{(i)}$, defined as the expected number of secondary infections that one infectious host can produce during its lifetime as an infectious host placed in an entirely susceptible population, leading to the threshold condition for an epidemic outbreak in the population, $R_0^{(i)}>1$~\cite{Anderson1992}. 
In our study, we assume that the two strains have the same basic reproductive number, $R_0^{(1)}=R_0^{(2)}=R_0$, but different infectious periods. This represents a case in which the pathogens have the same transmission potential and generate outbreaks characterized by the same impact on the population expressed in terms of attack rates. The epidemic waves are however different, unfolding with different timescales, the faster being the one characterized by the shorter generation time. In our case the generation time is uniquely determined by the infectious period~\cite{Anderson1992} (see Figure~\ref{fig:model}) and without  loss of generality we consider a \emph {fast} strain with infectious period $\mu_f^{-1} \equiv \mu^{-1}$ and a \emph{slow} strain with infectious period $\mu_s^{-1} \equiv \tau \mu^{-1}$. The parameter  $\tau>1$ quantifies  the timescale separation. 

The infection transmission is modeled by dividing the population of $N$ individuals into four compartments: susceptible $(S)$, infected by the fast strain $(I_f)$, infected by the slow strain $(I_s)$ and recovered $(R)$, i.e. immune to both strains. Each susceptible individual can contract either strain with the corresponding force of infection, $\beta_s I_s/N$ or $\beta_f I_f/N$, where $\beta_s$ $(\beta_f)$ is imposed by the equivalence of the basic reproductive numbers of the two strains. The two infection events are independent and mutually exclusive because of the assumptions considered. 

\subsection*{Host metapopulation network}

The multipathogen disease dynamics affects a spatially structured population of hosts modeled through a metapopulation system. This theoretical framework was first used in population ecology, genetics and adaptive evolution to describe population dynamics whenever the spatial structure of populations is known to play a key role in the system's evolution~\cite{Bascompte1998,Tilman1997,Hanski1997,Hanski2004} and later applied to understand the epidemic dynamics on such substrates~\cite{Anderson1984,May1984,Bolker1993,Lloyd1996,Keeling2002}. For the case of epidemic modelling, the infectious disease spreads in an environment characterized by a non-continuous spatial distribution of susceptible hosts and the pathogen diffusion depends on the ability of hosts to move from one region of the system to another one, connecting otherwise isolated communities~\cite{Hanski1997, Hanski2004}. 
Hosts mix homogeneously within the local communities (also called subpopulations or patches or nodes of the metapopulation network), whereas at the global, system-wide level, patches are coupled through the migration of hosts (represented in terms of mobility connections between patches), as schematically shown in Figure~\ref{fig:model}.

Here we consider a metapopulation network with $V= 10^4$ subpopulations. To each node $i$, we assign an initial number of individuals, $N_i$, and a degree $k_i$ denoting the number of connections the node has with other subpopulations in terms of mobility processes. The degrees of the nodes are distributed according to a given probability distribution $P(k)$, which we choose to represent the two most abundant situations in real systems~--~ namely, a Poisson distribution accounting for homogeneous networks of contacts and a power-law functional form which represents the case in which mobility patterns are highly heterogeneous.  To compare the effects of changes in the structural pattern of the subpopulations with no variations of the corresponding average values, we set the average degree of both networks, $\bar k$, to be the same. Homogeneous networks are generated following the Erd\H{o}s-R\'{e}nyi algorithm~\cite{Erdos1959}, which consists of assigning a link between each pair of nodes with probability $\bar k/(V - 1)$. It models a fairly homogenous system with low degree fluctuations. On the other hand, heterogeneous networks characterized by a power-law degree distribution, $P (k) \sim k^{-\gamma}$ where we consider $\gamma=  2.2$, are generated using an uncorrelated configuration model~\cite{Molloy1998, Catanzaro2005}.  In this second case, the  probability distribution is characterized by a   second moment that is much larger than the first, which makes it critical to explicitly take into account degree fluctuations. This feature is distinctive of a vast majority of social and demographic systems that have been empirically characterized~\cite{Keeling2010,Bajardi2011b, Chowell2003, Barrat2004, Guimera2005, Brockmann2006, Gonzalez2008,Barrat2008}. 

\subsection*{Host mobility pattern}

In case of homogenous traveling probability, mobility fluxes are modeled by assigning to each individual in subpopulation $i$ a probability $p_i$ per unit of time to travel to another neighboring subpopulation $j$. We assume that such probability is constant across nodes, namely $p_i \equiv p$, and that individuals leaving a subpopulation $i$ choose at random one of the available $k_i$ links \cite{Colizza2008}, so that the probability of traveling from $i$ to $j$ is given by $p/k_i$. According to the value of $p$, different mobility scenarios emerge: high values of $p$ yield large mobility fluxes resulting in a well mixed metapopulation system where individuals easily move from one patch to another; on the contrary small probability values result in a dynamically fragmented scenario in which patches are fairly isolated. The mobility process is described by the following diffusion equation:
\begin{equation}
\partial_t N_i(t)= -p N_i(t) + \sum_{j \in \upsilon(i)} \frac{p}{k_j} N_j(t)\,
\label{eq:diffusion}
\end{equation} 
where the sum runs over the set $\upsilon(i)$ of the $i\textrm{' s}$ nearest neighbors. According to this equation the population distribution at equilibrium is given by
\begin{equation}
 N_i= k_i \bar N / \bar k,
 \label{eq:eqpop}
 \end{equation}
where $\bar N$ is the average population size.  The main variables used in the model and the corresponding ranges of values considered are reported in Table~\ref{tab:param}. We also tested in the Text S1 more realistic definitions of host mobility, following empirical findings.

\subsection*{Computational modeling of competing pathogens}

To simulate the spread of the two strains on the metapopulation system of susceptible hosts, we initialize the number of individuals of each subpopulation at the equilibrium value given by Eq.~(\ref{eq:eqpop}). We then seed $50$ randomly chosen subpopulations for each strain by setting a proportion equal to $0.1 \%$ of the local population size in the corresponding infectious class. These conditions ensure the start of the outbreak for each strain for the values of the basic reproductive number considered, and at the same time they aim to avoid competition at the initial stage of the multistrain epidemic. Values of the number of initially infected nodes different from 50 where tested in order to check that this does not alter the simulation results.
Once the system is initialized, the transmission dynamics of the two strains is reproduced by means of  Monte Carlo numerical simulations at the discrete individual level. We consider hosts as integer units and we explicitly simulate both their mobility among different subpopulations and the infection transmission within each subpopulation as discrete-time stochastic processes, with fixed time step representing the unitary time scale $\Delta t=1$ of the process. To this end, at each time step, the number of hosts traveling along any connection of the system belonging to any compartment and the number of new infectious and recovered hosts for each subpopulation are extracted randomly from binomial and multinomial distributions to consider all possible outcomes of these events. Further details on the algorithm used for the simulations, as well as initial conditions and parameters, are described in Section~1 of the Text S1.

For each set of parameters we simulate 2,000 stochastic realizations of the spatial epidemic spreading averaging over different initial conditions, and over different instances of the metapopulation network that defines the spatial structure of the system population. For each scenario, we collect statistics of epidemiological quantities, including the number of subpopulations affected by each strain, the outbreak probability, and the incidence and attack rates of each strain, both at the global level and within each subpopulation. This allows to monitor the evolution of the two epidemics, their impact on the system, and the result of the competition process. 

\subsection*{Invasion threshold of a pathogen in metapopulation models}

Several works have recently studied the global spreading of a single strain SIR-like epidemic in metapopulation models~\cite{Colizza2007a, Colizza2007b, Colizza2008, Balcan2011, Balcan2012, Meloni2011, Belik2011a, Belik2011b,Poletto2012, Apolloni2013}. The threshold condition $R_0>1$  is sufficient for an epidemic outbreak to occur in a given subpopulation, but it does not guarantee the disease is able to spread globally. Low diffusion rates may indeed hinder a pathogen to disperse to other patches before it goes extinct locally, thus preventing the persistence of the virus and its spatial spread in the host population. The global spreading  of an infectious disease in a  metapopulation model is captured by the definition of an additional predictor of the disease dynamics, $R_*$, regulating the number of subpopulations that become infected from a single initially infected subpopulation, analogously to the reproductive number $R_0$ at the individual level~\cite{Ball1997, Cross2005, Cross2007}. The parameter $R_*$ defines a global invasion threshold: the condition $R_*>1$ guarantees that the epidemic taking place in the seeding subpopulation is able to spread at the global scale reaching a non-infinitesimal fraction of the metapopulation system. $R_*$ depends on several factors, including disease parameters, demography, metapopulation network structure, travel fluxes and mobility timescales. 

Theoretical studies in \cite{Colizza2007a, Colizza2007b, Colizza2008, Balcan2011, Balcan2012, Poletto2012, Apolloni2013, Belik2011a, Belik2011b,Meloni2011} have addressed the impact of empirically observed features on $R_*$, thus providing a better understanding about how mobility patterns and demography affect the invasion threshold of an infection. The current analytical framework allows to get an expression for $R_*$, in which the impact of several sources of heterogeneities in the topology of the metapopulation system, traffic fluxes \cite{Colizza2007a, Colizza2007b, Colizza2008, Balcan2011, Balcan2012,Meloni2011,Belik2011a,Belik2011b} and time scales \cite{Poletto2012} can be quantitatively assessed. In order to provide an understanding of the mechanisms shaping the global invasion condition of a multistrain epidemic, we review here the derivation of $R_*$ for the simplest case, of a single strain on a homogeneous metapopulation network with uniform mobility pattern. 

In a homogenous system, in which topological fluctuations can be neglected, all nodes can be assumed to have the same degree $\bar k$. If the mobility dynamics is described by Eq.~(\ref{eq:diffusion}), an expression for $R_*$ can be obtained by formalizing the seeding process of infected hosts through their migration from one patch to another. The probability $\mathcal P$ that an infected patch $i$ will seed the epidemic in a disease-free patch $j$ is given by $\mathcal P= 1-\left ( \frac{1}{R_0}\right )^{\lambda_{ij}}$ \cite{Bailey1975}, where $\lambda_{ij}$ is the number of infectious hosts who traveled from $i$ to $j$ during the entire duration of the outbreak, while infectious. The latter quantity can be estimated as follows. The total number of individuals that experience the disease during the epidemic unfolding within the subpopulation $i$ will be $\alpha N_i$, where $\alpha$ is the attack rate given by the SIR equations and $N_i$ is equal to $\bar N$ for all nodes -- as recovered by Eq.~(\ref{eq:eqpop}) in the case $k_i\equiv \bar  k$. Each infected individual stays in the infectious state for an average time $\mu^{-1}$ equal to the inverse of the recovery rate, during which it can travel to the neighboring subpopulation at rate $p/\bar k$. To a first approximation we can therefore consider that the number of  seeds sent from $i$ to $j$ during the duration of the outbreak is given by $\lambda_{ij}= \frac{p\alpha \bar N}{\mu \bar k}$. If we model the invasion from one patch to another in terms of a branching process, we obtain  that an infected subpopulation infects on average $(\bar k-1)\mathcal P$ subpopulations, where $\bar k-1$ is the number of connections along which the disease can spread. This leads to the following expression for $R_*$ in the homogeneous assumption 
\begin{equation}
R_{*}= \left(\bar k-1\right)\left(1-\left(\frac{1}{R_{0}}\right)^{\frac{p\alpha \bar N}{\mu \bar k}}\right).
\label{eq:thresh_hom}
\end{equation} 

As discussed, the global invasion threshold $R_*$ quantifies the spreading potential of an epidemic at the global level. For any set of parameters values characterizing the infection dynamics, the threshold condition $R_*>1$ defines a critical value of the host mobility below which the epidemic is not able to spread globally. It is worth remarking that this transition cannot be uncovered by continuous deterministic models because of the stochastic features of the contagion process and the discrete nature of circulating hosts.

Let us now consider the case of two competing strains~--~one slow and another fast. Even if both strains have the same transmission potential at the local level, namely the same $R_0$, their large scale spreading potential, encoded in $R_*$, would be different. As shown by Eq.~(\ref{eq:thresh_hom}), $R_*$ is indeed an increasing function of the infectious period, therefore $R_*(\mu_f)<R_*(\mu_s)$. This indicates that, in a metapopulation system of fully susceptible hosts, the slow strain would be able to infect on average a larger number of subpopulations than the fast strain, although at a much slower pace. As we will see in the following section the trade-off between transmission potential and spreading time-scale crucially impacts  the population level competition among the two epidemics. 

% Results and Discussion can be combined.

%.................. RESULTS

\section*{Results} \label{sec:results}

\subsection*{Two-strain competition}

We consider two strains with relatively high transmission potential, i.e. $R_0= 1.8$, and infectious rates given by $\mu=0.6$ and $\tau= 2$. As an indicator of the outcome of the competition between the two strains we consider the final number of subpopulations $D_\infty^s$ and $D_\infty^f$ affected by each strain during the outbreak. We say that a patch has been affected by a strain if at least a fraction $\alpha_T$ of the population within the patch has contracted the disease. We set $\alpha_T$ equal to $10\%$ and we checked that the results are not sensitive to the value of this parameter. By looking at the average of $D_\infty^s$ and $D_\infty^f$ when $p$ varies, we inspect several competition scenarios that are determined by mobility regimes. 

\begin{figure}[!ht]
\begin{center}
\includegraphics{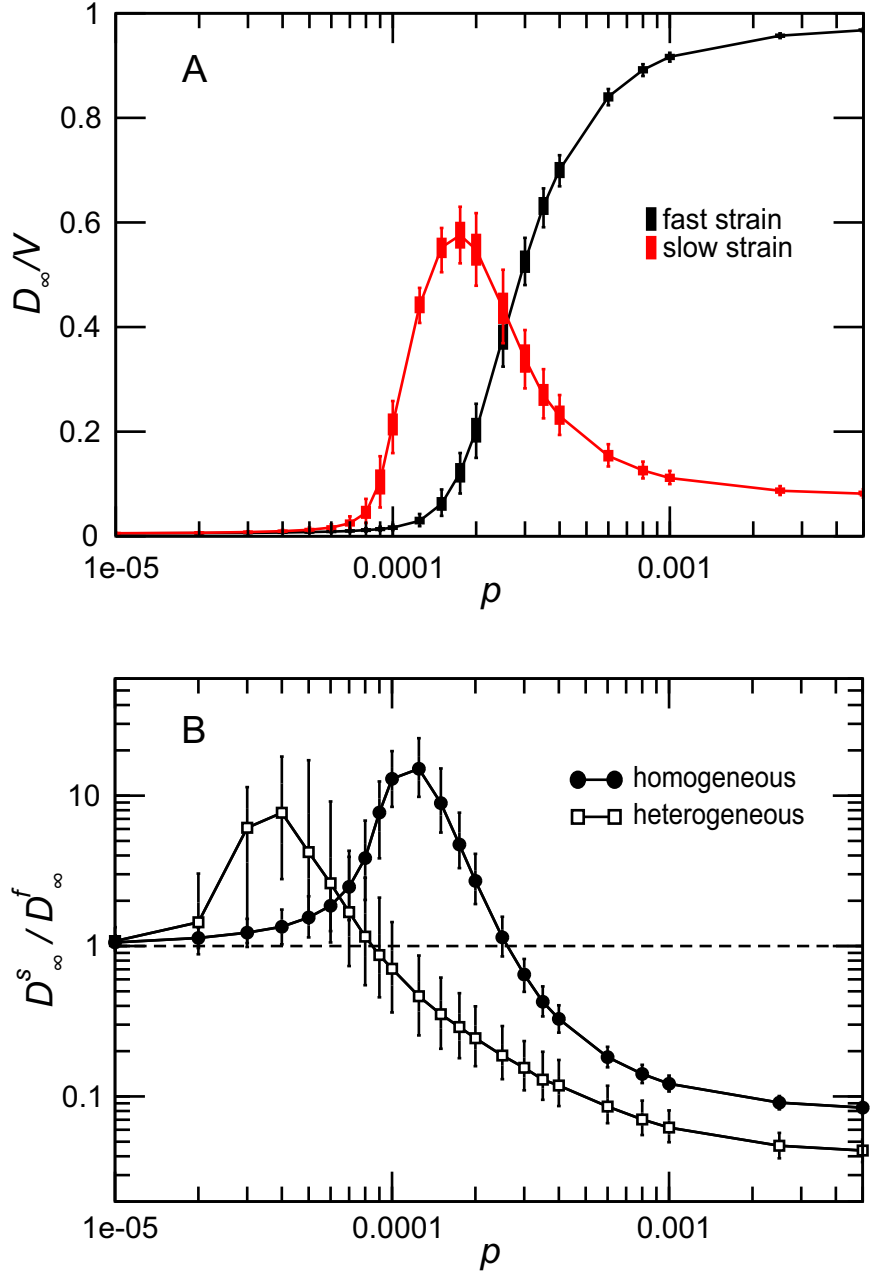}
\end{center}
\caption{
{\bf Competition between strains.} (A) Fraction of subpopulations infected by the fast and slow strains as a function of $p$. The quantity plotted is the median and the $95\%$ confidence interval over 2000 stochastic runs. Simulations were performed on a random homogeneous network. (B) Ratio  $D_\infty^s/D_\infty^f$ as a function of $p$ for both homogenous and heterogenous networks. The inter-quartile range is not displayed for the sake of visualization. In both panels the networks have average degree $\bar k= 5$. Both strains have $R_0= 1.8$.  Other parameters are $\mu= 0.6$ and $\tau= 2$. 
}
\label{fig:main_result}
\end{figure}

Figure~\ref{fig:main_result}A shows the results of the multistrain epidemic simulations assuming a homogeneous metapopulation structure. Different mobility regimes give rise to different coexistence and dominance patterns. For large values of $p$ the fast strain dominates affecting the vast majority of subpopulations infected in the system, the slow strain being constrained to  roughly  $\sim 10\%$ of the patches. As the value of $p$ decreases, the system-wide spreading potential of the slow strain progressively grows at the expense of the fast one, until a cross-over takes place at diffusion rate $p_c$. This intermediate  regime is characterized by the codominance of the two strains~\cite{Finkelman2007}, each one affecting approximately the same portion ($\sim 40\%$)  of infected subpopulations. Below this point, the slow strain becomes dominant, whereas  the fast one only induces local outbreaks propagating through a small number of subpopulations. Eventually, for very small values of $p$ none of the strains is able to spread geographically and no global outbreak occurs. Figure~\ref{fig:main_result}b further illustrates this  phenomenology by plotting the average value of the ratio  $D_\infty^s/D_\infty^f$ as a function of $p$. Values of the ratio much larger than 1 indicate the dominance of the slow strain, and values corresponding to $D_\infty^s/D_\infty^f \ll 1$ to the opposite scenario in which the fast strain dominates. The codominance phase is obtained for values of the ratio $D_\infty^s/D_\infty^f$ close to 1, with the  cross-over diffusion rate $p_c$ given by the intersect of the curve with  the horizontal line $D_\infty^s/D_\infty^f \equiv 1$. The figure also compares heterogeneous and homogeneous metapopulation systems. The results show that the behavior is qualitatively the same for both network structures,  the main quantitative difference being given by a lower $p_c$ value in the heterogeneous case. Similar results are also recovered with a different model for the mobility fluxes as detailed in Section~2 of the Text S1. 

The observed behavior can be understood according to the following intuitive explanation. After the two epidemics are seeded in their initial locations they evolve independently at the beginning, until one of the two strains reaches a subpopulation that has already been infected by the other strain, thus finding part of the population immune, i.e. a reduced pool of susceptible hosts to infect. This may prevent the strain to widely spread within the patch and diffuse further along the mobility connections towards other neighboring nodes. This competing mechanism favors the strain that spreads more rapidly and more efficiently from one patch to another, features that change depending on the mobility regime. When the traveling rate is high, the  whole system is at risk of a major epidemic because of the large rate of mixing across different patches.  Both $R_*^s$ and $R_*^f$  are much greater than 1, implying that the two epidemics would successfully reach the global invasion of the system, in absence of competition. When the two strains are competing on the same metapopulation system, the relevant factor for dominating the spread is given by the spreading speed; the shorter the infectious period and the more rapidly the strain reaches a large fraction of the system patches that will thus not be invaded by the slow strain. However by decreasing the value of $p$, $R_*$ of each strain also decreases. In the proximity of the invasion threshold the condition $R_*^s > R_*^f$ becomes relevant for the spreading dynamics and favours the slow strain which percolates more efficiently through the network. Indeed, the global epidemic time scale is not anymore dominated by the local velocity of transmission but rather by the mobility time scale of individuals. Hosts contracting the slow strain remain infectious for a longer time and thus have more chances to migrate while infectious. The low mobility rate, coupled with a short infectious period, hinders the movement of infectious hosts, resulting in a lower probability $\mathcal P$ of infecting a neighboring patch. We provide  a more quantitative understanding of the crossover behavior in the section dedicated to the analytical discussion of the results.

The same argument applies to explain the difference between the two network topologies observed in Figure~\ref{fig:main_result}B. The topological fluctuations that characterize the heterogeneous topology induce larger values of the parameter $R_*$ with respect to the corresponding homogeneous network (provided that the rest of parameters is kept the same)~\cite{Colizza2007a, Colizza2007b, Colizza2008}. Therefore the invasion threshold $R_*$ becomes larger than one for the two strains for smaller values of the mobility rate in the heterogenous case, which results in a shift of the cross-over diffusion rate $p_c$ towards lower values.

In the case of partial cross-immunity presented in the Text S1, we find that the main results reported for the full cross-immunity scenario still hold. Specifically, we have simulated situations in which recovered individuals from one strain may have up to $80\%$ cross-immunity to the other strain, which roughly correspond to estimates for diverse degrees of antigenic drift of influenza \cite{Koelle2006}.

\subsection*{Within-patch coexistence and spreading pattern}

\begin{figure}[!ht]
\begin{center}
\includegraphics{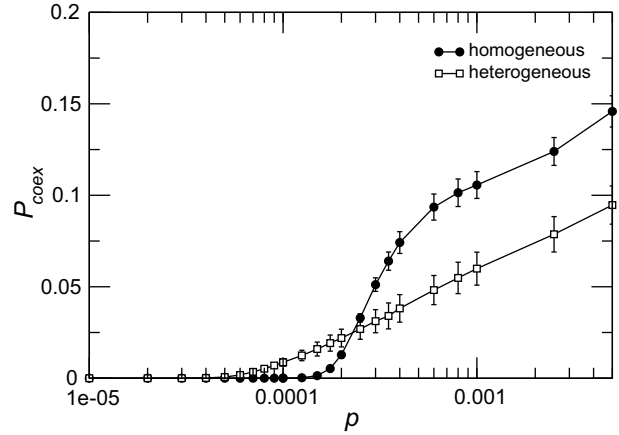}
\end{center}
\caption{
{\bf Coexistence probability within the same patch.} $P_{\textrm{coex}}$ is defined as the probability that within the same subpopulation both strains produce at least $1\%$ of the population infected. The quantity plotted is the average and the standard deviation over 2000 runs. The parameters used for the simulations are the same as in Figure~\ref{fig:main_result}.
}
\label{fig:coexistence} 
\end{figure}

We now focus on characterizing the coexistence of both strains at the within-patch level and their spread at the global spatial level. We define the coexistence probability $P_{\textrm{coex}}$ as the probability that within the same subpopulation both strains produce at least 1\% of the population infected. For both heterogeneous and homogeneous mobility networks $P_{\textrm{coex}}$ is an increasing function of the traveling rate $p$ (Figure~\ref{fig:coexistence}), therefore mobility favors the coexistence of the two strains within the same subpopulation. Coexistence is however generally unlikely to occur in a vast fraction of subpopulations, given the relatively small values of the probability obtained, showing that the two strains rarely coexist within the same subpopulation and the competition takes place at the metapopulation level.	

\begin{figure}[!ht]
\begin{center}
\includegraphics{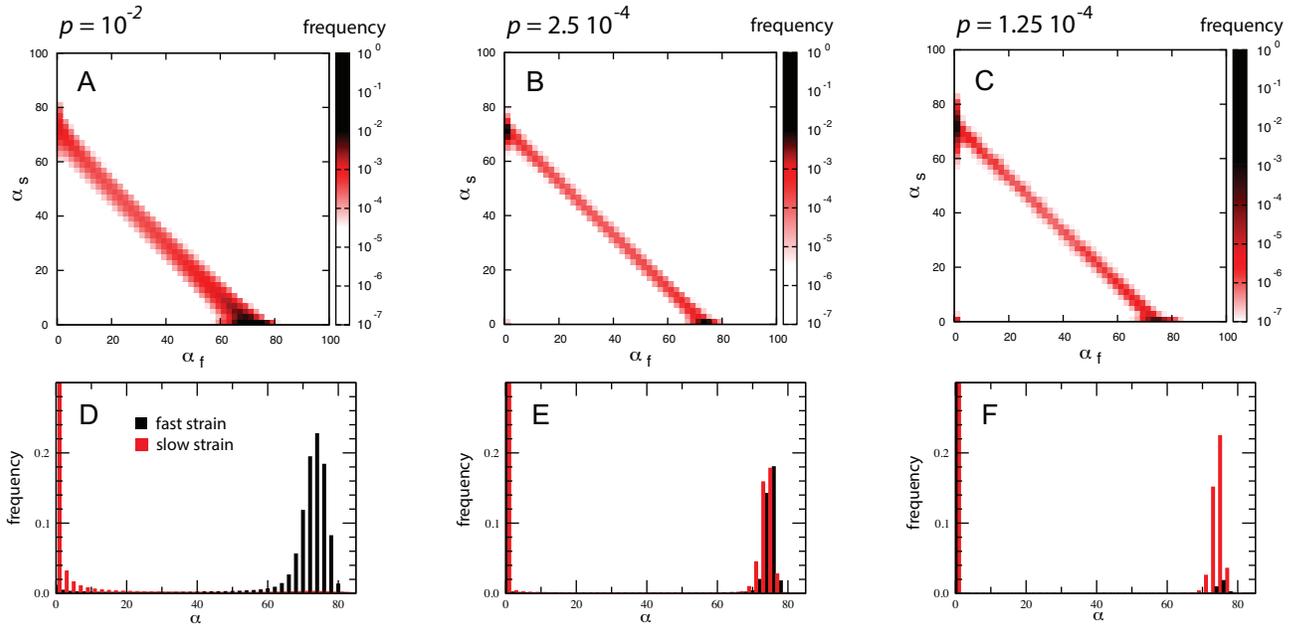}
\end{center}
\caption{
{\bf Within-patch coexistence and strain-specific attack rates.} A,B,C: heatmaps showing the frequency of occurrence of a given epidemic outcome $(\alpha^f,\alpha^s)$ within the patches, expressed in $\%$ as obtained by numerical simulations. D,E,F: histogram of the within-patch attack rate $\alpha$ (in $\%$) for the slow and fast strains. From left to right, three different mobility regimes are displayed: $p= 10^{-2}$ in which the fast strain dominates (A,D), $p= 2.5 \cdot 10^{-4}$ corresponding to the cross-over point (B,E), and $p= 1.25 \cdot 10^{-4}$ in which the slow strain dominates (C,F).
}
\label{fig:coexistence_in detail}
\end{figure}

To further characterize the two strain coexistence within a patch, we measure for each patch $i$ the attack rate at the end of the outbreak, defined by a two-dimensional variable $(\alpha_i^f,\alpha_i^s)$, where $\alpha_i^f\textrm{ }(\alpha_i^s)$ is the fraction of hosts affected by the fast (slow) strain within the patch during the outbreak. In all mobility regimes explored, the interaction between the two strains can be mapped to a small region of the $(\alpha^f,\alpha^s)$ space. Specifically, the strains always produce attack rates with a strictly linear dependence (Figure~\ref{fig:coexistence_in detail}, panels A,B,C), characterized by probability  distributions centered around
$\alpha=0$ and $ \alpha=75\%$  and with different proportions in the three diffusion regimes considered (Figure~\ref{fig:coexistence_in detail}, panels D,E,F). Moreover configurations in which only one strain is present in a subpopulation have a frequency of occurrence much higher than  configurations where the two strains co-exist, further confirming the results of Figure~\ref{fig:coexistence}.

\begin{figure}[!ht]
\begin{center}
\includegraphics{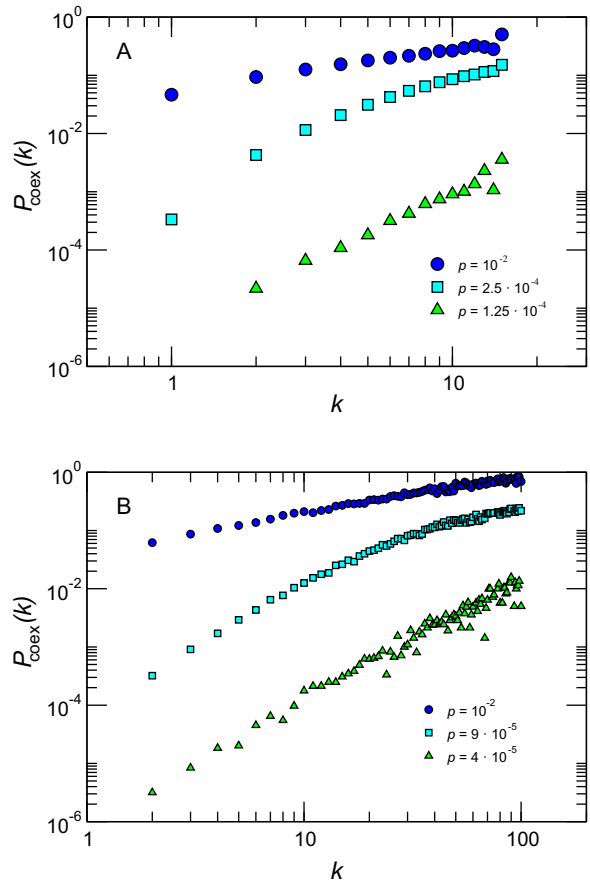}
\end{center}
\caption{
{\bf Probability of coexistence within a patch as a function of the patch connectivity} \mbox{ \boldmath $k$}.  Homogenous (A) and heterogeneous (B) cases are shown. Different traveling regimes are compared: they correspond to the scenarios in which the fast strain dominates (the highest value of $p$ considered in the two plots), the two strains coexist (intermediate value of $p$) and the slow strain dominates (smallest value of $p$). The quantity plotted is the average over 2000 runs; error bars are not displayed for the sake of visualization.
}
\label{fig:degree_block}
\end{figure}

We explore whether the coexistence of the two strains at the local level may carry a spatial signature. In absence of georeferenced data in our model that is based on an abstract spatial network, we consider the topological properties of the patches as possible spatial indicators. Noticeable differences arise when the probability of within-patch coexistence is measures by degree classes (Figure~\ref{fig:degree_block}). In both homogenous (panel A) and heterogenous (panel B) cases, $P_{\textrm{coex}}(k)$ is an increasing function of $k$ and it can vary over more than two orders of magnitude from poorly connected subpopulations to most connected ones. This behavior, although expected because highly connected patches are more likely to collect individuals from other subpopulations, highlights two different levels for strains competition in the system. 
On one side, in highly connected nodes the two strains compete at the single subpopulation level and the predominance of one of the two strains is dictated only by their epidemic parameters. Such behavior is mostly due to the fact that highly connected nodes are almost surely reached by infected individuals of both strains at the early stage of the spreading process. Thus, both strains are likely to infect a non-vanishing fraction of the node population at the same time, leading to higher probability of coexistence. On the other hand, as low connected nodes are harder to reach, the competition is mostly driven by the time at which one strain reaches the subpopulation.  The first strain to disseminate to the low connected patch has likely enough time to infect a large fraction of the susceptible hosts before the arrival of the other strain. In this case the competition between the two strains acts at the metapopulation level as coexistence between the strains is almost zero. 

\begin{figure}[!ht]
\begin{center}
\includegraphics{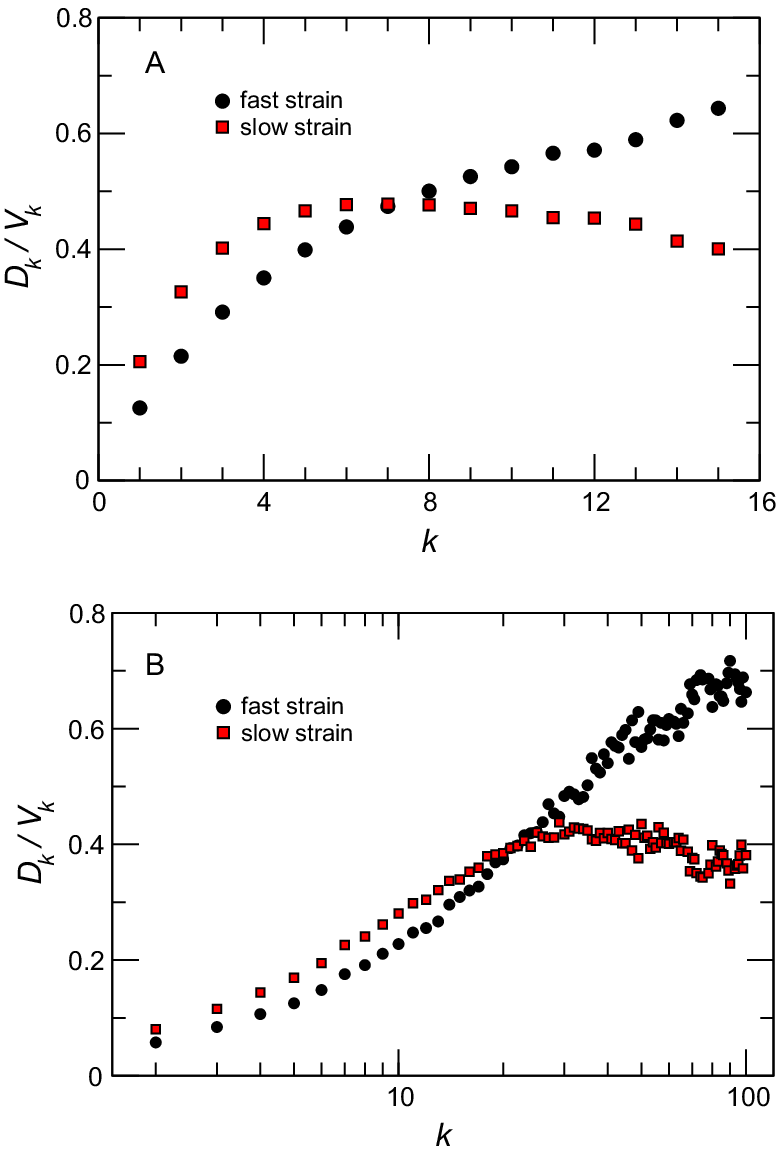}
\end{center}
\caption{
 {\bf Competition between strains per connectivity class in the cross-over regime.} Fraction of subpopulations infected by each strain within the degree class $k$, $D_k/ V_k$,  in the homogeneous (A) and heterogenous (B) networks. The two plots depict the behavior in the cross-over mobility regime ($p \simeq 2.5 \cdot 10^{-4}$ in  panel (A) and $p \simeq 8 \cdot 10^{-5}$ in panel (B)). The quantity plotted is the average over 2000 runs; error bars are not displayed for the sake of visualization.
}
\label{fig:attack_degree_block} 
\end{figure}

To conclude our analysis of the system at the patches' degree level in Figure~\ref{fig:attack_degree_block} we present the fraction of infected subpopulations $D_k/V_k$ with degree $k$ for the two strains as a function of the degree in the cross-over mobility region. In both homogenous and heterogenous networks the slow strain shows a higher incidence for low-degree nodes, whereas for intermediate and higher connectivities, the fast strain dominates the spreading process.

\subsection*{Impact of $R_0$ and $\tau$}

\begin{figure}[!ht]
\begin{center}
\includegraphics{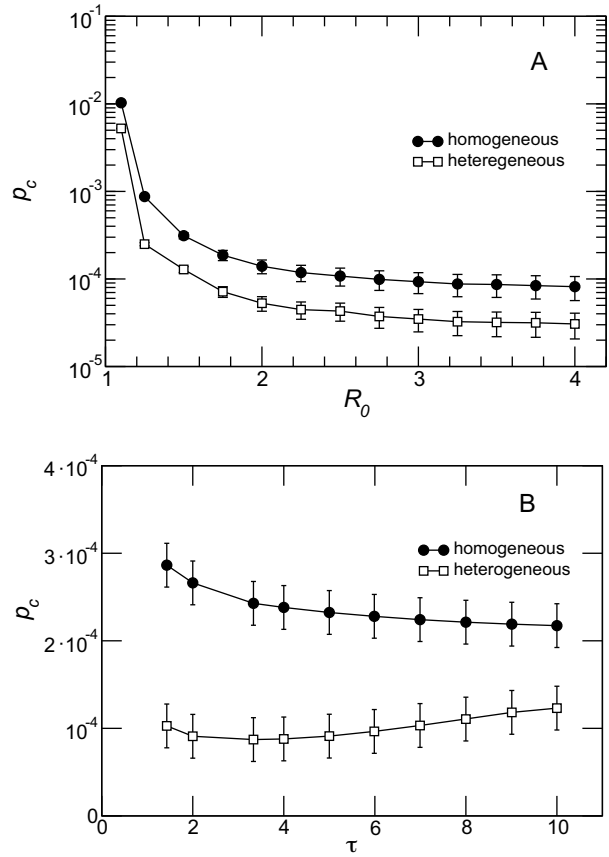}
\end{center}
\caption{
{\bf Dependence of the cross-over diffusion rate on the epidemiological parameters.} Cross-over diffusion rate  $p_c$  along with estimation error as a function of the reproductive number $R_0$  (A) and of $\tau$ (B) in the homogeneous and heterogeneous cases. The networks have average degree $\bar k= 5$. Other parameters are $\mu= 0.6$, $\tau= 2$  (A) and $R_0= 1.8$ (B).
}
 \label{fig:pc_varying_parameters}
\end{figure}

Finally,  we focus on the two parameters that  mainly affect the spreading of the strains and their interaction~--~namely, the reproductive number $R_0$ of both strains and the ratio $\tau$ between the infectious periods of the two strains. Variations of $R_0$ from 1.1 to 4 induce a variation of the cross-over mobility rate $p_c$ of more than two orders of magnitude (Figure~\ref{fig:pc_varying_parameters}A), with higher values of the basic reproductive number leading to smaller values of $p_c$. The decrease observed in the cross-over rate is very rapid for $R_0<2$, followed then by an almost constant value for larger values of $R_0$ in both network types, indicating the presence of  a critical $R_0$ beyond which the interaction dynamics of the two strains is dominated only by the disease parameters and not by the mobility rate. 

Differently from $R_0$, variations of $\tau$ do not strongly alter the value of the cross-over mobility rate, with a change of $\tau$ of one order of magnitude inducing variations in $p_c$ of less than $20\%$ (Figure~\ref{fig:pc_varying_parameters}B). Moreover, the initial fall off observed for $R_0$ at fixed $\tau$ (panel a) is not seen anymore. 
In both plots we note that the critical diffusion rate is smaller in the heterogeneous networks with respect to the homogenous ones for the whole range of parameters explored, confirming a favoring effect in the spatial spread of both strains as previously discussed.

To provide a specific example, we applied this framework to the case of two influenza-like strains spatially circulating on the real worldwide aviation network (assuming full cross-immunity and epidemiological parameters as in Figure~\ref{fig:pc_varying_parameters}B), we obtain that the air-transportation mobility scenario falls in the regime in which the fast strain is dominant for all the values of $\tau$ tested (more details are reported in the Text S1).

\subsection*{Analytical discussion} \label{sec:th_understand}

Here we focus on the case of homogeneous networks and propose a simplified analytical description of the dynamics to gain theoretical insights to further support the observed numerical behavior. We consider a continuous time approximation and assume that two strains do not interact at the early stage of the spreading process, in order to provide an estimation of the critical diffusion rate $p_c$ below which we have the dominance of the slow strain  and above which we have the dominance of the fast strain. The basic approach is to treat the dynamics at the system level in terms of the usual SIR model in a well mixed population, considering the subpopulations as the elementary ingredients of the spreading process. Under this assumption, the number of infected subpopulations $D(t)$ grows exponentially in time, and we can write
\begin{equation}
D(t) \sim e^{\frac{1}{T}\left(R_{*}-1\right)t},
\label{eq:cont}
\end{equation}  
where $T(\mu,\bar N,R_{0})$ is the duration of the outbreak in a single population and $R_*$ is the estimator of the invasion potential, as described in the Methods section, i.e. is the analogous of the basic reproductive number $R_0$ at the metapopulation level. 

If we consider the case of  two epidemics starting at different seeded subpopulations, by  neglecting possible interactions among the two strains, we obtain that the ratio between the number of subpopulations infected by the slow strain $(D_s)$ and the number infected by the fast one $(D_f)$ is given by:
\begin{equation}
\frac{D_{s}(t)}{D_{f}(t)} \sim  e^ {\left(\frac{\left(R^s_{*}-1\right)}{T_{s}} - \frac{\left(R^{f}_{*} - 1\right)}{T_{f}}\right) t}.
\label{eq:2strain_ratio}
\end{equation}   
Our goal is to derive the cross-over diffusion rate $p_c$ at which we have that both strains cocirculate, which is given by the condition $\frac{D_{s}(t)}{D_{f}(t)}=1$. Hence, from Eq.~(\ref{eq:2strain_ratio}), we get
\begin{equation}
\frac{R^s_{*}-1}{T_{s}} - \frac{R^{f}_{*} - 1}{T_{f}}= 0.
\label{eq:co_cond}
\end{equation}  
In the case of equal size populations, and for the same $R_0$, it is possible to show that the timescale defining the epidemic unfolding, for instance the maximum of the removal rate, is well approximated by a linear dependence on $\mu^{-1}$~\cite{ Bailey1975,Murray2005}. We can therefore assume that $T_s=T\tau$ and substituting Eq.~(\ref{eq:thresh_hom})  into Eq.~(\ref{eq:co_cond}) we explicitly arrive to the crossover condition as
 \begin{eqnarray}
\frac{1}{\tau} \left[\left(\bar k-1\right)\left(1-\left(\frac{1}{R_{0}}\right)^{\frac{p_c\alpha \bar N \tau}{\mu \bar k}}\right)-1 \right]= \nonumber \\
=\left(\bar k-1\right)\left(1-\left(\frac{1}{R_{0}}\right)^{\frac{p_c\alpha \bar N}{\mu \bar k}}\right)-1,
\label{eq:cross-over}
\end{eqnarray} 
where $T$ simplifies and disappears from the equation.

\begin{figure}[!ht]
\begin{center}
\includegraphics{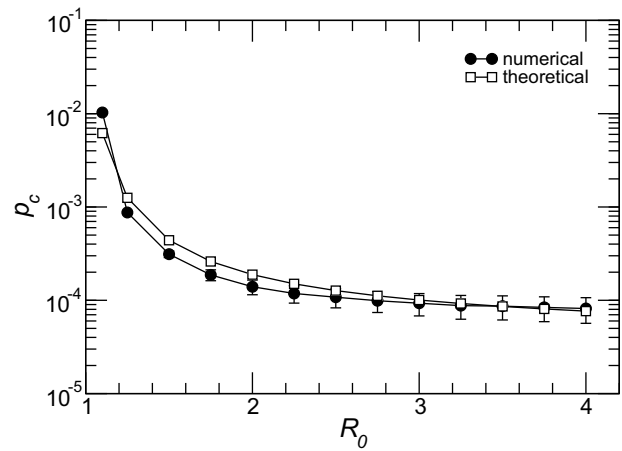}
\end{center}
\caption{
{\bf Theoretical predictions.} Comparison between the numerical and theoretical cross-over diffusion rate $p_c$  as a function of the reproductive number $R_0$ for the case of homogeneous network. Numerical results are the average over $2000$ stochastic runs, whereas theoretical values are obtained solving Eq.~(\ref{eq:pc}). The networks have average degree $\bar k= 5$. Other parameters are $\mu= 0.6$ and $\tau= 2$.
}
\label{fig:th_num}  
\end{figure}

Finally, denoting:
\begin{equation}
\left(\frac{1}{R_{0}}\right)^{\frac{p_c\alpha \bar N}{\mu \bar k}} = x,
\label{eq:x}
\end{equation}
 we have: 
 \begin{equation}
\left(\bar k-1\right)\left(1-x^{\tau}\right)-1= \tau \left(\left(\bar k-1\right)\left(1-x\right)-1\right).
\label{eq:pc}
\end{equation} 
Eq.~(\ref{eq:pc}) can always be solved for $p_c$ numerically and, in some cases, analytically. The comparison between theoretical predictions and numerical simulation results shows a good agreement in the behavior of the cross-over diffusion rate as a function of $R_0$ (Figure~\ref{fig:th_num}), confirming that the analytical approximation is able to capture the fundamental mechanisms for competition between the two strains.

%.................. DISCUSSION

\section*{Discussion}

\label{sec:con}
We studied a two-pathogen interaction in a spatially structured population of susceptible hosts mediated by immunological mechanisms (full cross-immunity) and ecological ones (hosts mobility),
where other biological and epidemiological features are kept equal across pathogens (basic reproductive number). Assuming the two diseases to be imported locally in different patches, we find that a variety of scenarios  emerge as a result of the competition between pathogens, driven by the host mobility rate. Either both infectious agents cocirculate and codominate  in the system, each of them reaching a substantial fraction of the patches, or one of the two dominates constraining the other to a rapid extinction. The  spatial structure enables the selection for a given trait depending on the hosts behavioral ecology regarding mobility. A longer infectious period constitutes a disadvantage for a rapidly mixing population across different patches as it generates a slower epidemic at the local level and therefore a slower invasion at the spatial scale. If the typical timescale for host mobility increases, the longer period during which hosts remain infectious make the invasion process more efficient with respect to the faster strain. 

We  found that in all cases the two strains rarely coexist within the same patch. Therefore, the competition occurs at the metapopulation level and it is determined by the spreading pattern at large spatial scales which in turn depends on the structure of the mobility network. Several works have recently shown the crucial role of host dispersal in mediating multi-strain interaction and in canalizing the evolution of pathogens traits \cite{Haraguchi2000, Ballegooijen2004, Wild2009, Boots2010,Boerlijst2010}. Our model contributes to this research efforts by focusing on the specific aspect of infectious duration and providing a clear understanding of how the interplay between the time scales of the dynamical processes involved~--~the unfolding dynamics of the two epidemics and host mobility dynamics~--~affects multi-strain competition. Therefore it highlights a mechanism that plays a  potentially relevant role on the process of pathogen evolution. Moreover, given that strains can only interact when they coexist, our results are of further interest as they show under what mobility conditions this interaction at the subpopulation level is feasible.

Our results show that there exist a codominance regime around the cross-over host diffusion rate $p_c$, where each infectious agent accounts for a proportion approximately equal to $40\%$ of the subpopulations of the system. However dominance of a single strain is more likely to occur than codominance as an outcome of competition, as measured by the larger interval in the phase space corresponding to a strain invading the majority of the patches. This result is consistent with the laboratory confirmed influenza surveillance data in the Northern and Southern hemisphere showing that H1 and H3 subtypes are rarely found in the same season in a given country (1 out of 171 country influenza seasons analyzed)~\cite{Finkelman2007}.

In the model we considered full cross-immunity among the circulating strains, a situation applicable, e.g., to measles infections, characterized by complex recurrent epidemics arising from cyclic exhaustion of susceptible hosts in the population~\cite{Grenfell2001}. This assumption is also often considered as a simplification when modeling multiple strains of influenza, though immunity after infection is strain-dependent and only partial cross-immunity against viral variants is found~\cite{Webster1992}. 
We have  explored situations of partial  cross-immunity showing that our findings are stable for relatively high degrees of cross-immunity between the two strains considered. These results thus show that our framework may  be applicable to two strains having a high level of similarity in their genetic and antigenic profiles, as this provides large cross-immunity across influenza strains. A full exploration of the spectrum of cross-immunity values is needed to further investigate to what extent they may affect the findings of this work. 

The model may also be extended to more than two interacting pathogens. While straightforward from a design point of view, increasing the number of pathogens rapidly increases the complexity of the system and the corresponding computational time of its numerical simulations, so that targeted methods need to be developed to reduce the exponentially large state spaces~\cite{Kryazhimskiy2007}. 

Furthermore our model considered an infection dynamics acting on timescales  much shorter than the host lifetime, and no demographic processes were therefore taken into account. In order to study outbreaks on longer timescales or that occur in recurrent cycles, mechanisms for susceptible hosts replenishments in the population need to be considered, as for instance birth and death processes  in the case of measles epidemics or loss of immunity in the case of influenza infection. This latter case would correspond to a two-strain SIRS compartmental approach and it could be used within our framework to study the role of host mobility on strain replacement events, as it may occur after influenza pandemics, where we need to assume that the other strain is already present and at equilibrium when an additional strain emerges in the system. While an application to human influenza A seems plausible with the limitations discussed above, a more comprehensive understanding of the general evolutionary dynamics of influenza viruses, central to its surveillance and control, would need to include punctuated antigenic change~\cite{Smith2004},  reassortment events~\cite{Holmes2005,Lindstrom2004,Nelson2006}, multiple circulating lineages~\cite{Holmes2005}, among other factors. 

The simplicity of the approach, on the other hand, allows us to provide analytical insights and theoretical predictions that further support the numerical results obtained with mechanistic discrete stochastic simulations. Such predictions are obtained with a very simplified mathematical reasoning, and here we discuss the main assumptions considered. We assumed that the two epidemics do not interact at the early stage, which is strictly verified only in the limit of infinite network size. Moreover, in using SIR-like equations for the dynamics of the number $D(t)$ of infected patches, treated as a continuous variable in the continuous time approximation, we  supposed that the infectivity of a node decays exponentially over time. However, in general, the infectivity of a subpopulation is proportional to the number of infectious individuals present in that subpopulation, which has a more complex functional dependence on $t$. Notwithstanding these approximations, the theoretical estimates for the cross-over diffusion rate $p_c$ are in good agreement with the values recovered numerically, for a large range of $R_0$ values. 

It is also worth remarking that the presented framework  is valid not only for human mobility and human multistrain epidemics, but it also applies to farmed or wild animals for which data on movements are available or can be partially mapped, along with the corresponding virological and serological data. The model could be for instance considered to investigate the role of bovine displacements among premises in a given country~\cite{Keeling2010,Bajardi2011b} and for import/export  across countries in the competition among foot-and-mouth disease strains~\cite{Haydon2001} following episodic invasion events~\cite{Keeling2005,Green2006}, or in the cocirculation of new serotypes of bluetongue virus following importation in Europe in 2006 and 2007~\cite{Saegerman2008}. Changing dynamics of dominant serotypes of rabies viral infections may be also related to changes in hosts movements (induced e.g. by changes in the local environment or ecosystem disturbances), in addition to other mechanisms~\cite{Bourhy2005}. Variations in  hosts behavioral ecology may be tested to further investigate the interactions among multiple subtypes of avian influenza virus in specific settings, given their importance in the possible occurrence of reassortment events leading to the emergence of novel viruses~\cite{Monne2012}. Here we focused specifically on directly transmitted diseases that can be well described by the homogeneous mixing assumption within each local community of hosts coupled by spatial propagation due to host migrations among communities. 

\section*{Supporting Information}
{\bf Text S1}. The File contains the details on the model implementation, the study of the scenarios with heterogenous mobility patters  and with partial cross-immunity and the discussion of the framing of the study in a realistic case. 
% Do NOT remove this, even if you are not including acknowledgments

%\section*{References}

% \bibliography{template}

\vspace{7cm}

\section*{Figure Legends}


\vspace{0.5cm}

\section*{Tables}

\begin{table*}[!ht]
\begin{tabular}{|l|l|l|}
\hline
{\bf Variable} & {\bf Description} & {\bf Values}\\
\hline
$V$ & number of patches & $10^4$\\
$\bar N$ & average host population size per patch & $10^4$ \\
$k$ & patch degree, i.e. number of connections to other patches & average value $\bar k=5$\\
$P(k)$ & degree distribution & homogenenous (Poisson)  \\
& & or heterogeneous $(P(k)\simeq k^{-2.2})$\\
$R_0$ & reproductive number assumed to be equal across strains & $[1.1-4]$\\
$\tau$ & scaling factor of slow strain's infectious period & $[1.5-10]$\\
& to fast strain's infectious period & \\ 
$p$ & uniform probability of hosts migration & $[10^{-5}-10^{-2}]$\\
\hline
\end{tabular}
\caption{ \label{tab:param} Model variables and their corresponding values. }
\end{table*}

\end{document}